\documentclass[aps,pra,twocolumn,showpacs,floatfix,tightenlines,superscriptaddress]{revtex4}
\usepackage{graphicx,bm}
\usepackage{amssymb}

\begin{document}

\newcommand{\proofend}{\hfill\fbox\\\smallskip }

\title{Measurement-induced Nonlinearity in Linear Optics}
\author{Stefan Scheel}\email{s.scheel@imperial.ac.uk}
\affiliation{Quantum Optics and Laser Science, Blackett Laboratory,
Imperial College London, Prince Consort Road, London SW7 2BW, United
Kingdom}

\author{Kae Nemoto}
\affiliation{School of Informatics, Dean Street, Bangor University,
Bangor LL57 1UT, United Kingdom} 
\affiliation{Hewlett Packard Laboratories, Filton Road, Stoke 
Gifford, Bristol BS34 8QZ, United Kingdom}

\author{William J. Munro}
\affiliation{Hewlett Packard Laboratories, Filton Road, Stoke 
Gifford, Bristol BS34 8QZ, United Kingdom}
\affiliation{Quantum Optics and Laser Science, Blackett Laboratory,
Imperial College London, Prince Consort Road, London SW7 2BW, United
Kingdom}

\author{Peter L. Knight}
\affiliation{Quantum Optics and Laser Science, Blackett Laboratory,
Imperial College London, Prince Consort Road, London SW7 2BW, United
Kingdom}
\date{\today}

\begin{abstract}
We investigate the generation of nonlinear
operators with single photon sources, linear optical elements and appropriate 
measurements of auxiliary modes. We provide a framework for the 
construction of useful single-mode and two-mode quantum gates necessary 
for all-optical quantum information processing. We focus our 
attention generally on using minimal physical resources while providing a
transparent and algorithmic way of constructing these operators.
\end{abstract}

\pacs{03.67.-a, 42.50.-p, 03.67.Lx, 03.65.Ta}

\maketitle
%%%%%%%%%%%%%%%%%%%%%%%%%%%%%%%%%%%%%%%%%%%%%%%%%%%%%%%%%%%%%%%%%%%%%%
\section{Introduction}

\begin{figure}[ht]
\centerline{\includegraphics[width=6cm]{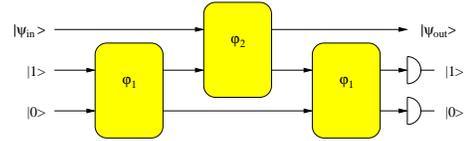}}
\caption{\label{fig:KLM} Schematic setup of the KLM circuit for 
generating a nonlinear sign shift using 3 beam-splitters, a single photon 
source and single photon resolving detectors.}
\end{figure}

In recent years we have the seen signs of a new technological revolution 
in information processing, a revolution caused by a paradigm shift to
information processing using the laws of quantum
physics \cite{dowling02}. Since the pioneering  work of Feynman
\cite{feymann82}, Deutsch \cite{deutsch85}, and Shor \cite{shor95}  
a significant effort has occurred worldwide to 
develop the tools necessary to realise such a revolution. There are many 
possible routes and architectures \cite{machines,NielsenChuang}
available to develop these quantum information  
processing devices. It has long been thought that photons would be 
an extremely strong contender for realising some quantum information 
processing circuits \cite{milburn88}. Many of the photon's properties,
for instance easy manipulation,  
have made them ideal for this. However, for scalable quantum
information processing  
we require photons to interact with one another. To achieve such
interactions it was  
known that massive reversible nonlinearities would be required
\cite{shen84}. Materials giving  
such large nonlinearities were thought to be (and are still) well
beyond our ability  
to manufacture. Knill, Laflamme, and Milburn (KLM) however found a way to 
create such nonlinearities using only linear optical elements, single photon 
sources and detectors \cite{KLM}. More precisely they showed how it is
possible  
using such elements to  perform conditionally the nonlinear transformation 
\begin{eqnarray}
|\Psi\rangle&=& c_0 |0\rangle+c_1 |1\rangle+c_2 |2\rangle \nonumber \\
&\rightarrow& c_0 |0\rangle+c_1 |1\rangle-c_2 |2\rangle=|\Psi'\rangle \,.
\end{eqnarray}
The optical circuit (depicted in Fig.~\ref{fig:KLM}) creating this nonlinear 
transformation uses ancilla modes, one prepared with a single photon present 
and the other empty. The nonlinearity was induced by definite measurements of 
the presence of the single photon and the vacuum state in the
appropriate ancilla modes.  
This insight has reopened the door to all-optical quantum information
processing.   
Other optical schemes \cite{footnote-1}
have been proposed along the
KLM line to generate such sign shifts 
%\cite{ralph02,Ralph01,pittman01,rudolph01,ralph02a,kok02,munro03,koasho01,knill01}.
\cite{ralph02a,Ralph01,kok02,others}.
These operations 
are generally conditional in nature. By this we mean the
transformation only works when  
the appropriate measurement results are obtained at the ancilla
detectors. While this would seem to limit  
the viability of the information processing, it is straightforward 
however by using a teleportation-based protocol to turn such
nondeterministic operations into  
deterministic ones \cite{gottesman99,KLM}. 

There have been a number of key experiments demonstrating elements of
linear optical  
information processing \cite{pittman02,pittman02a,pittman03}. These
have generally focused  
on the technology necessary to perform single-qubit rotations and CNOT
gates. Such gates  
are well known to be sufficient to perform universal computation 
(they are the minimum set required). From these primitive 
elements interesting devices such as Quantum Repeaters \cite{kok02a} and 
single-photon quantum non-demolition detectors \cite{kok02} can be created. 
In this paper we wish to shift the focus slightly. Instead of using only 
these primitive gates we will investigate what operations can be
constructed from  
linear elements, single photon sources, and detectors. This shift is
analogous to the  
shift in classical computing from a RISC (reduced instruction set)
architecture  
to the CISC (complex instruction set) architecture. The RISC-based
architecture in quantum computing  
terms could be thought of as a device built only from the minimum set
of gates while  
the CISC-based machine would be built from a much larger set, a
natural set of gates allowed by the  
fundamental resources.  

Our primary focus in this paper will be on the operations that can be
constructed  
from the linear optics set. We show how to construct general operators
that can be applied  
to the required input states. We further indicate what operations are
{\it easily} constructed and  
what are potentially difficult, illustrating our constructive
procedure with examples from  
one-mode and two-mode situations. Our constructive procedure can easily be
applied to multiple modes.  
The inputs to the computational modes do not need to be restricted to
qubits only:  
the operations can be applied onto qudits and continuous variables
just as easily. 

This paper is organised as follows. In Sec.~\ref{sec:general} we will
derive some general expressions necessary for the construction of
useful nonlinear operations. In Sec.~\ref{sec:single} we will be
concerned with single-mode operations, followed by two-mode operations
in Sec.~\ref{sec:twomode}. Until then, we assume perfect beam
splitters and detections which is an oversimplification, indeed. We
will therefore focus on the effects of absorption and non-unit
detection efficiencies in Sec.~\ref{sec:lossy} before drawing some
conclusions in Sec.~\ref{sec:conclusions}. Some useful formulae
regarding permanents of unitary matrices can be found in the Appendix.

%%%%%%%%%%%%%%%%%%%%%%%%%%%%%%%%%%%%%%%%%%%%%%%%%%%%%%%%%%%%%%%%%%%%%%
\section{General beam-splitter transformation}
\label{sec:general}

In order to introduce the notation we will be using throughout the paper we
will briefly review the most basic features of quantum-state
transformation by a lossless beam splitter. We refer the reader to the
extensive literature for 
details \cite{beamsplitter}. Every (lossless) beam splitter can be
thought of as a 
unitary operator on the level of photonic creation and annihilation
operators of the incoming and outgoing fields, i.e.
\begin{equation}
\hat{\bm{b}}
= \hat{U}^\dagger \hat{\bm{a}} \hat{U}
= \bm{\Lambda} \hat{\bm{a}} \,,\quad
\hat{\bm{a}}={\hat{a}_1 \choose \hat{a}_2}\,,\quad
\hat{\bm{b}}={\hat{b}_1 \choose \hat{b}_2}\,,
\end{equation}
where $\hat{U}$ is a unitary operator and $\bm{\Lambda}$ the
associated unitary matrix [$\bm{\Lambda}\in$ SU(2)]. The transformation
matrix $\bm{\Lambda}$ consists of the transmission and reflection
coefficients $T$ and $R$ and can be given in the form
\begin{equation}
\bm{\Lambda} = \left( \begin{array}{cc}
T & R \\ -R^\ast & T^\ast
\end{array} \right) \,.
\end{equation}
Unitarity of $\bm{\Lambda}$ requires $|T|^2+|R|^2=1$ which leads to
the usual definition of the beam splitter `angle' $\varphi$ by writing
$|T|=\cos\varphi$, $|R|=\sin\varphi$. The unitary operator $\hat{U}$ can
be given in several equivalent forms, two of which are the following:
\begin{eqnarray}
\hat{U} &=& e^{-i\hat{\bm{a}}^\dagger\bm{\Phi}\hat{\bm{a}}} \,,\quad
\bm{\Lambda}=e^{-i\bm{\Phi}} \,,\\
\label{su(2)}
\hat{U} &=& T^{\hat{n}_1} e^{-R^\ast \hat{a}_2^\dagger \hat{a}_1}
e^{R \hat{a}_1^\dagger \hat{a}_2} T^{-\hat{n}_2} \,.
\end{eqnarray}
The effect of the beam splitter cannot only be described by
transforming the photonic operators, but equivalently by transforming
the quantum state $\hat{\varrho}$ as
\begin{equation}
\hat{\varrho}_{\text{out}} = \hat{U} \hat{\varrho}_{\text{in}}
\hat{U}^\dagger \,.
\end{equation}
Noting that the input density operator $\hat{\varrho}_{\text{in}}$ can
be written as a functional of photonic creation and annihilation
operators,
$\hat{\varrho}_{\text{in}}=\hat{\varrho}_{\text{in}}[\hat{\bm{a}},\hat{\bm{a}}^\dagger]$,
the quantum-state transformation can be represented as
\begin{equation}
\hat{\varrho}_{\text{out}} =
\hat{\varrho}_{\text{in}}\left[\hat{U}\hat{\bm{a}}\hat{U}^\dagger ,
\hat{U}\hat{\bm{a}}^\dagger \hat{U}^\dagger\right] =
\hat{\varrho}_{\text{in}}\left[ \bm{\Lambda}^+ \hat{\bm{a}},
\bm{\Lambda}^T \hat{\bm{a}}^\dagger \right] \,,
\end{equation}
that is, the state transforms with the \textit{inverse} operator
\cite{plk1,stenholm1}. On 
the level of quantum states we thus have to perform the replacements
\begin{eqnarray}
\hat{\bm{a}} &\mapsto& \bm{\Lambda}^+ \hat{\bm{a}} \,,\\ \label{adagger}
\hat{\bm{a}}^\dagger &\mapsto& \bm{\Lambda}^T \hat{\bm{a}}^\dagger \,.
\end{eqnarray}
We will use Eq.~(\ref{adagger}) extensively throughout the paper.

Suppose we were given an input state with $N$ modes with the
associated creation and annihilation operators labelled by
$\hat{a}_i^{(\dagger)}$, $i=1\dots N$. Additionally, we have a supply
of $M$ auxiliary modes labelled by $\hat{a}_j^{(\dagger)}$,
$j=N+1,\ldots,N+M$. Then, a general unitary transformation on all the
modes maps $\hat{\bm{a}}^\dagger\mapsto\bm{\Lambda}^T\hat{\bm{a}}^\dagger$,
$\bm{\Lambda}\in\text{SU}(N+M)$. What we mean precisely by SU($N+M$) is
a unitary operator on the level of photonic creation and annihilation
operators in $N+M$ dimensions. In what follows, we will only make use
of the unitarity of the corresponding matrices and will not further
elaborate on the actual underlying group structure. In order to construct 
our quantum operations we will use the decomposition of an 
arbitrary element of the group SU($N$) into at most $N(N-1)/2$ U(2) 
group elements, i.e. beam splitters \cite{Reck94}.

First, let us define our notation. By $|0\rangle^{\otimes N}$ we mean the 
tensor product state $|0\rangle_1|0\rangle_2\ldots |0\rangle_N$.
Let the input state now be given in a functional form as
\begin{equation}
|\psi_{\text{in}}\rangle =
\hat{f}(\hat{a}_1^\dagger,\dots,\hat{a}_N^\dagger) |0\rangle^{\otimes N}
\end{equation}
and the auxiliary state in product form as
\begin{equation}
|\psi_{\text{aux}}\rangle = \prod\limits_{j=N+1}^{N+M}
\frac{\left(\hat{a}_j^\dagger\right)^{m_j}}{\sqrt{m_j!}}
|0\rangle^{\otimes M} \,.
\end{equation}
Here $m_j$ is a non-negative integer that represents the number of 
photons initially in the mode $j$.
Finally, the state we project on shall be denoted by
\begin{equation}
|\psi_{\text{proj}}\rangle = \prod\limits_{j=N+1}^{N+M}
\frac{\left(\hat{a}_j^\dagger\right)^{n_j}}{\sqrt{n_j!}}
|0\rangle^{\otimes M} \,.
\end{equation}
where $n_j$ represents the number of photons in the projected mode $j$.
The output state after mixing at the beam splitter network and projecting 
onto $|\psi_{\text{proj}}\rangle$  looks then as
\begin{widetext}
\begin{eqnarray}
|\psi_{\text{out}}\rangle &\propto& \langle \psi_{\text{proj}} | \hat{U}
|\psi_{\text{aux}}\rangle \otimes |\psi_{\text{in}}\rangle 
\nonumber \\
 &=& {}^{M \otimes} \langle 0| \prod\limits_{i,j=N+1}^{N+M}
\frac{\left(\hat{a}_i\right)^{n_i}}{\sqrt{n_i! m_j! }}
\left(\sum\limits_{k=1}^{N+M}\Lambda_{kj}\hat{a}_k^\dagger\right)^{m_j}
\hat{f}\left( \sum\limits_{l=1}^{N+M} \Lambda_{l1}\hat{a}_l^\dagger,\dots,
 \sum\limits_{l=1}^{N+M} \Lambda_{lN}\hat{a}_l^\dagger\right)
|0\rangle^{\otimes N+M} \,.
\end{eqnarray}
What we see here is that the effect of the beam splitter network is to
generate the desired mixing of the photonic creation operators of signal
and auxiliary modes.
Now we make use of the ordering formula well-known from bosonic operator 
algebras (see, e.g., \cite{VogelWelsch,Louisell})
\begin{equation}
\left[ \hat{a}, F(\hat{a},\hat{a}^\dagger) \right] =
\frac{\partial}{\partial \hat{a}^\dagger} F(\hat{a},\hat{a}^\dagger)
\end{equation}
to rewrite the output state as
\begin{eqnarray}
|\psi_{\text{out}}\rangle \propto 
{}^{M \otimes} \langle 0| \prod\limits_{i,j=N+1}^{N+M}
\frac{\left(\hat{a}_i\right)^{n_i}}{\sqrt{n_i! m_j!}}
\left(\sum\limits_{k=1}^{N+M}\Lambda_{kj}\hat{a}_k^\dagger\right)^{m_j}
\hat{f}\left( \sum\limits_{l=1}^{N+M} \Lambda_{l1}\hat{a}_l^\dagger,\dots,
 \sum\limits_{l=1}^{N+M} \Lambda_{lN}\hat{a}_l^\dagger\right)
|0\rangle^{\otimes N+M} \,.
\end{eqnarray}
\end{widetext}
Furthermore, we expand the function
$\hat{f}(\hat{a}_1^\dagger,\dots,\hat{a}_N^\dagger)$
in a Taylor series as
\begin{equation}
\hat{f}(\hat{a}_1^\dagger,\dots,\hat{a}_N^\dagger) =
\sum\limits_{p_1,\ldots,p_N=1}^N
c_{p_1,\ldots,p_N}
\frac{\left( \hat{a}_1^\dagger \right)^{p_1}}{\sqrt{p_1!}} \cdots
\frac{\left( \hat{a}_N^\dagger \right)^{p_N}}{\sqrt{p_N!}} \,.
\end{equation}
where $c_{p_1,\ldots,p_N}$ is constrained in such a way that 
$\sum\limits_{p_1,\ldots,p_N=1}^N |c_{p_1,\ldots,p_N}|^2=1$. 
In that way we obtain the action of a SU($N+M$)-network in a quite general 
way. In general, this can be a laborious task. In order to see the 
structure behind it, let us focus first onto single-mode signal states.
That is, the input state will be
\begin{eqnarray}
|\psi_{\text{in}}\rangle &=& \hat{f}(\hat{a}_1^\dagger) |0\rangle \\
&=& \sum\limits_m \frac{c_m}{\sqrt{m!}} \left( \hat{a}_1^\dagger
\right)^m | 0\rangle
\end{eqnarray}
and the network will represent an element of the group SU($N+1$).

\begin{figure}[!hbt]
\centerline{\includegraphics[width=4cm]{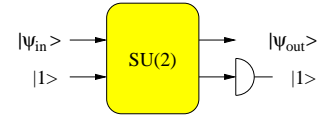}}
\caption{\label{fig:catalysis} Schematic setup for generating the
simplest non-unitary conditional operator with a single-photon input
and a single-photon detection.}
\end{figure}

In what follows we will restrict ourselves to the important special
case when our resources consist of single photons and single-photon
detectors. In this case, we can derive a number of interesting results.
Let us first start with a very simple (and in fact well-known) example, a 
single beam splitter.
Feeding a single photon in one input arm of the beam splitter and measuring 
a single photon leaving one output port of the beam splitter, we have in fact 
created the conditional non-unitary operator [using Eq.~(\ref{su(2)})]
\cite{CondMeas}
\begin{equation}
\hat{Y} = \langle 1_2 | \hat{U} | 1_2 \rangle =
T^{\hat{n}_1-1} \left[ |T|^2 -\hat{n}_1 |R|^2 \right]
\end{equation}
acting on some signal state $|\psi_{\text{in}}\rangle$ (see
Fig.~\ref{fig:catalysis}).
This is a very special result and probably the simplest non-unitary 
operator one can actually generate. This conditional operator has
already been realised in an experiment \cite{Lvovsky} where it is called
`quantum-optical catalysis'. 

In the following we will present some 
results on the general structure of conditional non-unitary operators.

\begin{figure}[!htb]
\begin{minipage}{4cm}
\includegraphics[width=3.8cm]{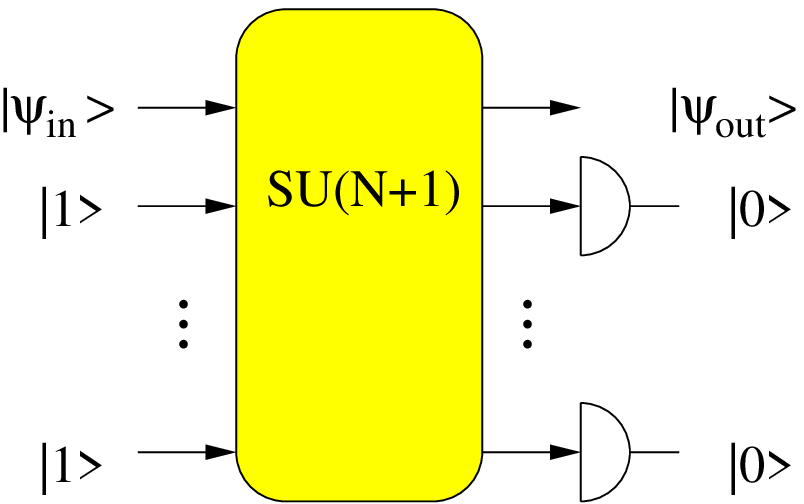}
\end{minipage}
\hfill
\begin{minipage}{4cm}
\includegraphics[width=3.8cm]{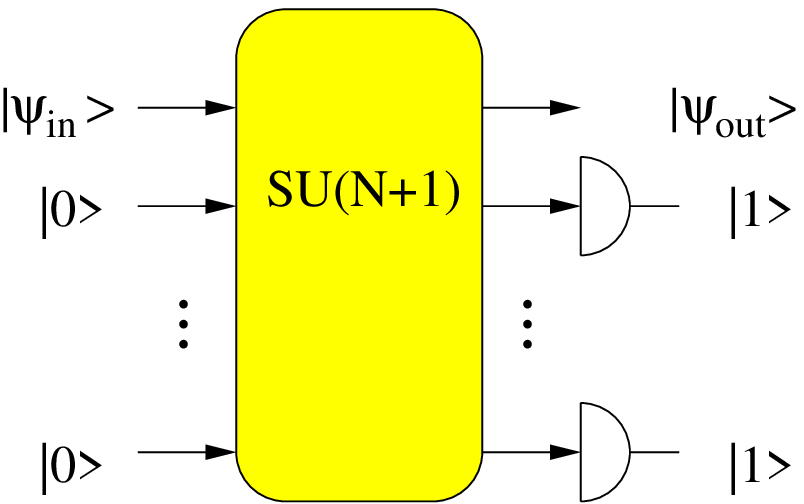}
\end{minipage}
\caption{\label{fig:th1} Adding (subtracting) photons to (from) the signal
mode by subtracting (adding) the corresponding number of photons from (to)
the auxiliary modes.}
\end{figure}

\vspace*{1ex}
\noindent
\textbf{Proposition 1}: Let us suppose all $N$ auxiliary modes are
prepared in single-photon states, and all $N$ detectors measure
vacuum. This is equivalent to acting with an operator
$\sim \left(\hat{a}_1^\dagger\right)^N$ on the signal state (left figure
in Fig.~\ref{fig:th1}).\\
\noindent
\textit{Proof}: 
The auxiliary and detected states are
\begin{equation}
|\psi_{\text{aux}}\rangle=\prod\limits_{i=2}^{N+1} \hat{a}_i^\dagger
|0\rangle^{\otimes N} \,,\quad
|\psi_{\text{det}}\rangle = |0\rangle^{\otimes N}\,.
\end{equation}
The conditional (un-normalised) output state is therefore
\begin{eqnarray}
|\psi_{\text{out}}\rangle &\propto&
\sum\limits_m \frac{c_m}{\sqrt{m!}}
\nonumber \\ && \hspace*{-5ex}
{}^{N \otimes}\langle 0| \left( \prod\limits_{i=2}^{N+1}
\sum\limits_{j=1}^{N+1} \Lambda_{ji} \hat{a}_j^\dagger \right)
\left( \sum\limits_{k=1}^{N+1} \Lambda_{k1} \hat{a}_k^\dagger
\right)^m | 0 \rangle^{\otimes N+1}
\nonumber \\ &=&
\sum\limits_m \frac{c_m}{\sqrt{m!}}
\left( \prod\limits_{i=2}^{N+1} \Lambda_{1i} \right) \Lambda_{11}^m
\left( \hat{a}_1^\dagger \right)^{m+N} |0\rangle
\nonumber \\ &=&
\left( \prod\limits_{i=2}^{N+1} \Lambda_{1i} \right)
\left( \hat{a}_1^\dagger \right)^N
\sum\limits_m \frac{c_m}{\sqrt{m!}} \Lambda_{11}^m
\left( \hat{a}_1^\dagger \right)^m |0\rangle
\nonumber \\ &=&
\left( \prod\limits_{i=2}^{N+1} \Lambda_{1i} \right)
\left( \hat{a}_1^\dagger \right)^N \Lambda_{11}^{\hat{n}_1}
|\psi_{\text{in}}\rangle \,.
\end{eqnarray}
Apart from normalisation (or success probability), which depends on
the chosen input state, the output state is proportional to the
$N$-fold application of the creation operator.\proofend

In complete analogy, we can prove the following:

\vspace*{1ex}
\noindent
\textbf{Proposition 2}: Let us suppose all $N$ auxiliary modes are
prepared in the vacuum state and each of the $N$ detectors measures a
single photon. Then, this is equivalent to acting with $\hat{a}_1^N$
on the input state (right figure in Fig.~\ref{fig:th1}).\\

\textit{Proof}: Again, let us first write down the auxiliary and the
detected state:
\begin{equation}
|\psi_{\text{aux}}\rangle=
|0\rangle^{\otimes N} \,,\quad |\psi_{\text{det}}\rangle =
\prod\limits_{i=2}^{N+1} \hat{a}_i^\dagger |0\rangle^{\otimes N}\,.
\end{equation}
Acting on the input state gives
\begin{eqnarray}
|\psi_{\text{out}}\rangle &\propto&
\sum\limits_m \frac{c_m}{\sqrt{m!}}
\nonumber \\ &&
{}^{N \otimes}\langle 0|
\left( \prod\limits_{i=2}^{N+1} \hat{a}_i \right)
\left( \sum\limits_{k=1}^{N+1} \Lambda_{k1} \hat{a}_k^\dagger
\right)^m | 0 \rangle^{\otimes N+1}
\nonumber \\ && \hspace*{-8ex} 
=\sum\limits_m \frac{c_m}{\sqrt{m!}}
{}^{\otimes N}\langle 0|
\left( \prod\limits_{i=2}^{N+1}
\frac{\partial}{\partial \hat{a}_i^\dagger} \right)
\left( \sum\limits_{k=1}^{N+1} \Lambda_{k1}  \hat{a}_k^\dagger
\right)^m | 0 \rangle^{\otimes N+1}
\nonumber \\ && \hspace*{-8ex}
= \sum\limits_m \frac{c_m}{\sqrt{m!}} \frac{m!}{(m-N)!}
\left( \prod\limits_{i=2}^{N+1} \Lambda_{i1} \right)
\Lambda_{11}^{m-N} \left( \hat{a}_1^\dagger \right)^{m-N} |0\rangle
\nonumber \\ && \hspace*{-8ex}
= \left( \prod\limits_{i=2}^{N+1} \Lambda_{i1} \right) 
\Lambda_{11}^{\hat{n}_1} \hat{a}_1^N |\psi_{\text{in}}\rangle \,,
\end{eqnarray}
where in the last line we have repeatedly made use of the formula
\begin{equation}
\left( \hat{a}^\dagger \right)^p |0\rangle = \frac{1}{p+1} \hat{a}
\left( \hat{a}^\dagger \right)^{p+1} |0 \rangle
\end{equation}
which immediately follows from the commutation relations of the
photonic operators. This proves that, indeed, measuring $N$ photons
from an $N$-mode auxiliary vacuum input is equivalent to acting
$N$ times with the annihilation operator on the signal state.\proofend

Propositions 1 and 2 show how to generate arbitrary powers of creation
and annihilation operators. In fact, one could have already guessed the 
general form of these operators by recalling that the network is 
represented by an element of the compact group SU($N+1$). Compactness of 
the group translates into photon-number conservation which is why adding 
(subtracting) $N$ photons from the auxiliary modes must end up as 
subtracting (adding) photons from (to) the signal mode.
Note that in both cases only the matrix elements $\Lambda_{i1}$ or 
$\Lambda_{1i}$ ($i=2,\ldots,N+1$),
respectively, appear. This means that the network decouples into a
sequence of $N$ disconnected beam splitters. That is already the
minimal number of beam splitters necessary for the generation of the
wanted operators.

The next step consists of showing how
powers of the number operator can be realised. In fact, an obvious way
would be to combine the results from Propositions 1 and 2 and
to construct an alternating network producing sufficient numbers of
creation and annihilation operators. This might not be the most sensible 
way to do. In fact, as we will see later, the following result has much 
stronger implications for the construction of interesting quantum 
operations. 

\begin{figure}[ht]
\centerline{\includegraphics[width=4cm]{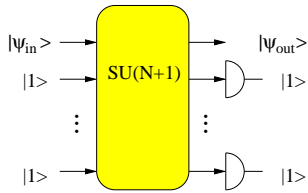}}
\caption{\label{fig:th2} Generating polynomials of photon-number operators
by single-photon inputs and detections.}
\end{figure}

\vspace*{1ex}
\noindent
\textbf{Proposition 3}: Measuring single photons in all $N$ detectors
from a supply of $N$ single-photon auxiliary state amounts to
multiplying the input state with a polynomial of $N$th degree in the
number operator, $P_N(\hat{n}_1)$ (Fig.~\ref{fig:th2}).\\
\textit{Proof}:
We will only sketch this proof and calculate the highest power of
$\hat{n}_1$ and leave the remaining terms for an interested reader to 
calculate. Given that we choose the auxiliary and detected states of the form
\begin{eqnarray}
|\psi_{\text{aux}}\rangle&=&\prod\limits_{i=2}^{N+1} \hat{a}_i^\dagger
|0\rangle^{\otimes N} \,,\nonumber \\
|\psi_{\text{det}}\rangle &=&
\prod\limits_{k=2}^{N+1} \hat{a}_j^\dagger |0\rangle^{\otimes N}\,,
\end{eqnarray}
the output state can be written in the following way:
\begin{widetext}
\begin{eqnarray}
\label{eq:prop3}
|\psi_{\text{out}}\rangle &\propto&
\sum\limits_m \frac{c_m}{\sqrt{m!}}
%\nonumber \\ &&
{}^{N \otimes}\langle 0| \left( \prod\limits_{k=2}^{N+1}
\frac{\partial}{\partial \hat{a}_k^\dagger} \right)
\left[ \prod\limits_{j=2}^{N+1} \left( \sum\limits_{i=1}^{N+1}
\Lambda_{ij} \hat{a}_i^\dagger \right) \right]
%\nonumber \\ &&
\left( \sum\limits_{n=1}^{N+1} \Lambda_{n1} \hat{a}_n^\dagger
\right)^m |0^{\otimes N+1}
\nonumber \\ &=&
\left( \prod\limits_{j=2}^{N+1} \Lambda_{1j} \right)
\left( \prod\limits_{n=2}^{N+1} \Lambda_{n1} \right)
\frac{\hat{n}_1!}{(\hat{n}_1-N)!} \Lambda_{11}^{\hat{n}_1-N}
|\psi_{\text{in}}\rangle
%\nonumber \\ &&
+\ldots +\left( \sum\limits_{j=2}^{N+1} \prod\limits_{i\in{\cal P}}
\Lambda_{ji_{\cal P}} \right) \Lambda_{11}^{\hat{n}_1}
|\psi_{\text{in}}\rangle \,.
\end{eqnarray}
\end{widetext}
In the first term the factorial $\hat{n}_1!/(\hat{n}_1-N)!$ is a
polynomial of order $N$ in $\hat{n}_1$ and thus the desired
result. All other terms (not written except for the last, in lowest
order in $\hat{n}_1$) contain lower-degree polynomials
\cite{footnote-2}. This proves the assertion.
\proofend

The simplest example of this proposition is a single beam splitter the 
result of which we have already seen in Eq.~(\ref{su(2)}). However, with 
the above propositions, we can immediately generalise our considerations 
to obtain the following results:
\begin{enumerate}
\item Given that the following for ancilla and detected modes
$$\begin{array}{rcl}
|\psi_{\text{aux}}\rangle &=& |1\rangle^{\otimes N+M} \\
|\psi_{\text{det}}\rangle &=& |1\rangle^{\otimes N} \otimes
|0\rangle^{\otimes M}
\end{array} \,,$$
the output state will be
$$|\psi_{\text{out}}\rangle \propto (\hat{a}_1^\dagger)^M P_N(\hat{n}_1)
|\psi_{\text{in}}\rangle \,.$$
We immediately see that this procedure has allowed us to act on the
input state with the creation operator $(\hat{a}_1^\dagger)^M$.
\item Analogously, with
$$\begin{array}{rcl}
|\psi_{\text{aux}}\rangle &=& |1\rangle^{\otimes N} \otimes
|0\rangle^{\otimes M} \\
|\psi_{\text{det}}\rangle &=& |1\rangle^{\otimes N+M} 
\end{array} \,,$$
the output state will be
$$|\psi_{\text{out}}\rangle \propto P_N(\hat{n}_1) (\hat{a}_1)^M 
|\psi_{\text{in}}\rangle \,.$$
\end{enumerate}
In both situations we have, with the aid of linear optics, single
photon sources and detectors, been able to operate on the input state
$|\psi_{\text{in}}\rangle$ with both $\hat{a}_1^M$ and
$(\hat{a}_1^\dagger)^M$. Let us now turn our attention to single-mode
operations that are of interest in connection with quantum information
processing.

%%%%%%%%%%%%%%%%%%%%%%%%%%%%%%%%%%%%%%%%%%%%%%%%%%%%%%%%%%%%%%%%%%%%%%
\section{Single-mode operations}
\label{sec:single}

From now on we will focus onto the generation of \textit{unitary}
operators which are of utmost importance for most quantum information
processing tasks. For all unitary operators it is easy to define the
success probability, since unitary operators leave the norm of a
quantum state unchanged. Since these operators $\hat{Y}$ are prepared
conditionally, the success probability is just
\begin{equation}
p_{\text{success}} = \| \hat{Y}|\psi\rangle \|^2
\end{equation}
for \textit{any} (normalised) state vector $|\psi\rangle$.

We can derive some interesting results about these unitary
operators. For example, let us suppose our input state is a
single-mode state consisting only of elements in the zeroth and first
Fock layer. It is clear that \textit{all} operations on
$|\psi_{\text{in}}\rangle$ of the type
\begin{equation}
|\psi_{\text{in}}\rangle = c_0|0\rangle +c_1|1\rangle \to
c_0 |0\rangle +e^{i\varphi} c_1|1\rangle
\end{equation}
can be realised with a probability of $p=1$, since unitary operations
simply consist of phase shifts of the $|1\rangle$ state. A special
example with $\varphi\!=\!\pi$ is the Pauli-$\hat{\sigma}_z$. Going
one step further we may ask what the conditions are for generation of
unitary operations on single-mode states with up to two photons. It is
reasonable to assume that we would need at least an SU(3)-network,
that is, two auxiliary modes.
In fact, we find that every unitary single-mode operator acting on
states with up to two photons, separately in each Fock layer, can be
generated by an SU(3)-network with two single-photon inputs and two
single-photon detections. In order to show that, let us first
calculate the conditional operator for the SU(3)-network with
$|\psi_{\text{aux}}\rangle=|\psi_{\text{det}}\rangle=|11\rangle$. We get
\begin{eqnarray}
\label{eq:unitary3}
\hat{Y}|\psi_{\text{in}}\rangle &=& \mbox{per } \bm{\Lambda}(1|1) |0\rangle
+\mbox{per }\bm{\Lambda} |1\rangle \nonumber \\ && \hspace*{-10ex}
+ \left( 2\Lambda_{11} \mbox{per }\bm{\Lambda} -\Lambda_{11}^2 \mbox{per }
\bm{\Lambda}(1|1) +2\Lambda_{12}\Lambda_{21}\Lambda_{13}\Lambda_{31}\right)
|2\rangle \,.\nonumber \\
\end{eqnarray}
It is known that the range of $\mbox{per }\bm{\Lambda}$ (as a function of
all its relevant parameter) is the unit disk in the complex plane
\cite{Minc} (see Appendix). In fact, so is the range of any principal 
sub-permanent $\mbox{per }\bm{\Lambda}(i|i)$. 
This can be seen from the decomposition of
an SU(3)-matrix in terms of a product of three SU(2)-matrices \cite{Reck94} 
which themselves have a range spanning the unit disk. Therefore, it is
immediately clear that we can again generate any phase
$e^{i\varphi_1}$ between the states $|0\rangle$ and $|1\rangle$. As
for the two-photon Fock layer, we can rewrite the coefficient in
Eq.~(\ref{eq:unitary3}) to obtain a condition on the matrix
$\bm{\Lambda}$ as
\begin{equation}
\label{eq:condition}
\mbox{per }\bm{\Lambda}(1|1) \left[ e^{i\varphi_2}
+\Lambda_{11}^2-2\Lambda_{11} e^{i\varphi_1} \right] =
2\Lambda_{12}\Lambda_{21}\Lambda_{13}\Lambda_{31} \,,
\end{equation}
where $e^{i\varphi_2}$ is the phase shift between $|0\rangle$ and
$|2\rangle$. The modulus of the rhs of Eq.~(\ref{eq:condition}) can be
shown to be bounded from above by $8/(27|\Lambda_{11}|^2)$ by noting
that $\prod_i \Lambda_{1i}$ is the product of the elements of a unit
vector. Noting also that the principal sub-permanent
$\mbox{per }\bm{\Lambda}(1|1)$ can take any value across the unit disk we
can conclude that Eq.~(\ref{eq:condition}) has 
always a solution. This in turn means that every unitary single-mode
operator acting within Fock layers on states with up to two photons 
can be generated by an SU(3)-network with two single-photon inputs 
and two single-photon detections which was to be proven. 
The probability of success is $|\mbox{per }\bm{\Lambda}(1|1)|^2$.
It is also possible however to create certain phase shifts with the necessity 
for two ancilla photons. For instance, in \cite{KLM} it was shown that
a sign shift on the $|2\rangle$ Fock state only is possible with the
ancilla state $|10\rangle$.

%%%%%%%%%%%%%%%%%%%%%%%%%%%%%%%%%%%%%%%%%%%%%%%%%%%%%%%%%%%%%%%%%%%%%%
\section{Two-mode operations}
\label{sec:twomode}

In order to do something useful in terms of quantum information
processing, we have to operate on two modes simultaneously. This can
be done in more than one way. For example, one can simply generalise
the theory presented above for a single signal mode to more than one
signal mode. It turns out that this is not a very transparent way. We
will follow another route instead and decompose the two-mode operation
into three subsequent steps:
\begin{enumerate}
\item combine the two modes at a beam splitter,
\item act on both modes \textit{separately},
\item and recombine the modes at another beam splitter.
\end{enumerate}
The effect of the beam splitters is to mix the modes and to make them
accessible for a \textit{single-mode} operation in such a way that we
can apply the result in Sec.~\ref{sec:single}.

\subsection{The `controlled-phase' gate}

We will illustrate this statement with an example. Consider the
two-mode operator $\hat{C}_\varphi$ acting on qubits. 
Its truth table is
\begin{eqnarray}
|00\rangle &\to& |00\rangle\,,\nonumber\\
|01\rangle &\to& |01\rangle\,,\nonumber\\
|10\rangle &\to& |10\rangle\,,\nonumber\\
|11\rangle &\to& e^{i\varphi}|11\rangle\,.
\end{eqnarray}
In terms of photon creation and annihilation operators the operator 
$\hat{C}_\varphi$ can be represented as
\begin{equation}
\label{eq:cphase}
\hat{C}_\varphi = 1-(1-e^{i\varphi})\hat{n}_1\hat{n}_2 \,.
\end{equation}
Now let us assume that we mix the signal modes at a symmetric beam
splitter. The operator $\hat{C}_\varphi$ acts only in
the two-photon Fock layer. 
Then it is very easy to see that with (nonlinear)
single-mode operators
$\hat{N}_i=1-\frac{1}{2}(1-e^{i\varphi})\hat{n}_i(\hat{n}_i-1)$, $i=1,2$, 
we achieve a transformation of an input state
\begin{equation}
|\psi_{\text{in}}\rangle = c_{00} |00\rangle + c_{01} |01\rangle +
c_{10} |10\rangle +c_{11} |11\rangle
\end{equation}
into
\begin{equation}
\hat{C}_\varphi |\psi_{\text{in}}\rangle =
c_{00} |00\rangle + c_{01} |01\rangle +
c_{10} |10\rangle +c_{11}e^{i\varphi} |11\rangle \,.
\end{equation}

\begin{figure}[ht]
\centerline{\includegraphics[width=8cm]{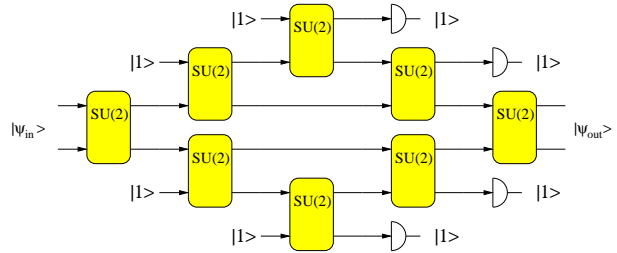}}
\caption{\label{fig:cphase1} Controlled-phase gate with single-photon
detectors only.}
\end{figure}

The nonlinear operator needed on both modes are polynomials of second
degree in the number operators $\hat{n}_i$ and can thus be prepared
conditionally with two auxiliary modes prepared in single-photon Fock
states on each side followed by double single-photon detection. Hence,
the overall requirements are four single-photon sources, eight beam
splitters, and four single-photon detectors. The generic network is
shown in Fig.~\ref{fig:cphase1}. The detectors all measure single
photons. We can write down the conditional operator as
\begin{eqnarray}
\hat{Y}|\psi_{\text{in}}\rangle &=& \mbox{per }\bm{\Lambda}(1|1)
c_0|0\rangle + \mbox{per }\bm{\Lambda} c_1|1\rangle
\nonumber \\ && \hspace*{-8ex}
+ \left[2\Lambda_{12}\Lambda_{21}\Lambda_{13}\Lambda_{31}
+2\mbox{per }\bm{\Lambda} -\Lambda_{11}^2 \mbox{per }\bm{\Lambda}(1|1)\right]
c_2|2\rangle \,.\nonumber \\
\end{eqnarray}
The success probability is $|\mbox{per }\bm{\Lambda}(1|1)|^2$. Numerically,
we find values up to $p_{\text{success}}\approx 0.24$ in each
interferometer arm.

\begin{figure}[ht]
\centerline{\includegraphics[width=8cm]{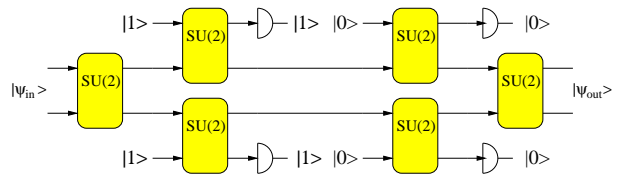}}
\caption{\label{fig:cphase2} Controlled-$\hat{\sigma}_z$ gate with
single-photon and vacuum detectors.}
\end{figure}

However, it turns out that there is an even simpler network
with only six beam splitters and two single-photon sources
\cite{Ralph01}. It has the 
disadvantage, though, that one needs two vacuum detectors which are
hard to make (and which are pretty inefficient). The corresponding 
network is shown in Fig.~\ref{fig:cphase2}.
The set of beam splitters fed with vacuum  states act as conditional phase
shifts. In summary, we find that the beam splitters must satisfy
\begin{eqnarray}
\arg T_{|1\rangle} &=& -\arg T_{|0\rangle} \,,\\
|T_{|1\rangle}| &=& 0.476 \,,\\
|T_{|0\rangle}| &=& 0.87 \,,
\end{eqnarray}
which gives a success probability of $p_{\text{success}}\approx 0.23$
in each arm, hence a total success probability of $\approx 0.05$.

Let us remark that the controlled-$\hat{\sigma}_z$ investigated by Ralph
\textit{et~al.} \cite{Ralph01} falls into the same category as that
described in our Fig.~\ref{fig:cphase1}. The difference is that one of
the single photons in each arm of the interferometer is replaced by
the vacuum state and the single-photon detector by a vacuum detector
\cite{vacuum-detector}, respectively. This network corresponds to the
following conditional operator:
\begin{eqnarray}
\hat{Y}|\psi_{\text{in}}\rangle &=& \Lambda_{22} c_0 |0\rangle
+\mbox{per }\bm{\Lambda}(3|3)c_1|1\rangle \nonumber \\ &&
+(2\Lambda_{12}\Lambda_{21}\Lambda_{11}+\Lambda_{22}\Lambda_{11}^2)
c_2|2\rangle \,.
\end{eqnarray}
The probability of success is $|\Lambda_{22}|^2$. One needs to satisfy
the set of conditions
\begin{eqnarray}
\label{eq:constraintsRalph}
\mbox{per }\bm{\Lambda}(3|3) &=& \Lambda_{22} \,,\\
2\Lambda_{12}\Lambda_{21}\Lambda_{11}+\Lambda_{22}\Lambda_{11}^2 &=&
-\Lambda_{22} \,,
\end{eqnarray}
from which it immediately follows that $\Lambda_{11}\!=\!1-\sqrt{2}$. The
maximal value $|\Lambda_{22}|^2$ can take under the constraints
(\ref{eq:constraintsRalph}) is then indeed 0.25 which is why the gate
in Ref.~\cite{Ralph01} is indeed optimal.

\subsection{The `swap' gate}

A somewhat more interesting operator is the swap operator $\hat{S}$ in
the sense that here we encounter the first example of an operator that
needs fewer resources than one would expect when considering CNOT and
single-qubit rotations as building blocks for quantum circuits.
It is known that it can be made from three CNOT operators
$\hat{\not{\hspace*{-0.5ex}C}}$ (equivalent to controlled-$\hat{\sigma}_z$ 
gates with attached Hadamard gates). Acting on qubits, one can write the
photonic-operator version of it as 
\begin{equation}
\hat{S} = \hat{n}_1\hat{n}_2 +(\hat{n}_1-1)(\hat{n}_2-1)
-\hat{a}_1^\dagger\hat{a}_2(\hat{n}_1-1)
-\hat{a}_2^\dagger\hat{a}_1(\hat{n}_2-1) \,.
\end{equation}
Let us see how the single-mode version of $\hat{S}$ can be derived. It
is immediately clear that we have to act on the single-photon Fock
layer only. It turns out that the nonlinear single-mode operators are
\begin{eqnarray}
\hat{N}_1 &=& 1+2\hat{n}_1(\hat{n}_1-2)\,,\\
\hat{N}_2 &=& 1 \,,
\end{eqnarray}
which means that we do nothing on mode 2, and we act with a polynomial
of second degree in $\hat{n}_1$ on mode 1. Therefore, we would need
only two single-photon sources, four beam splitters, and two
single-photon detectors. However, the operator $\hat{N}_1$, when
acting on Fock states $|n\rangle$, is nothing but a single-mode phase
shift $(-1)^{\hat{n}_1}$. That is, the whole network collapses into a
single $\pi$-phase plate in one arm of the Mach-Zehnder
interferometer leaving us with just two beam splitters and one phase
plate. This gate is remarkable in the sense that it is also
\textit{unconditional}, that is, it works \textit{deterministically}
with unit probability which makes it rather special.

These two simple examples show a general principle of constructing
these networks. Both operators have in common that they act only
within a specific Fock layer ($\hat{S}$: one photon;
$\hat{C}_\varphi$: two photons). One then projects out
all those Fock layers which are not affected by the operator. This
leads to the polynomials in the number operators. The design of the
polynomial coefficients in each case depend on the specific operation
one wants to achieve.

\subsection{General considerations}

A general conclusion can already be drawn from the results on one- and
two-qubit operators: It is highly desirable to rewrite the quantum
information network in such a way that the actual computation can be
made as long as possible in the same Fock layers. Every crossing to
another layer (cf. the Pauli operators $\hat{\sigma}_x$ and
$\hat{\sigma}_y$) requires additional resources which might not be
necessary. This leads us to state our main result of this paper:

\vspace*{1ex}
\noindent
\textbf{Theorem}:
The generic operations that can be done easily and effectively with linear 
optics are operations within the same Fock layers. Let $M$ be the number 
of signal modes we want to operate on. Any $M$-qubit gate acting within 
Fock layers can be constructed with the help of generalised 
Mach--Zehnder interferometers with $M$ input and output ports 
($2M$-ports for short) and at most $M$ conditional operators generating 
polynomials in the number operator of at most $M$th order (equivalent 
to SU($M+1$)-networks).\\
\textit{Proof}: 
The proof of this assertion is now straightforward. Any operator acting 
within Fock layers can be written as a polynomial of at most $M$th order 
in all photon number operators. The $2M$-port mixes all the $M$ input modes 
in such a way that we are left with a tensor product of $M$ operators 
inbetween the $2M$-ports conditionally generating polynomials of at most 
$M$th order in the individual photon number operators. \proofend

This result shows how to construct these operations in an algorithmic 
fashion. That is what we mean with `easy'. Since there is no inherent 
exponential scaling of the success probability with respect to the number 
of modes (qubits) we act on, there is a good reason to call them also 
`effective'.

Unfortunately, not all two-qubit gates can be written in terms of a
Mach-Zehnder interferometer and appropriate single-mode
operations. Perhaps the most notorious example is the CNOT
gate. Although similar to the controlled-$\hat{\sigma}_z$, there is no
way to find an interferometric setup that `disentangles' the two modes
in such a way that there existed single-mode operators that performed
the sought task. The proof of this statement goes along the following
lines: Let us call $\hat{U}(\varphi)$ the beam splitter operator that
rotates the qubit axes by an angle $\varphi$ [see Eq.~(\ref{su(2)}); a
Mach-Zehnder interferometer would consist of a succession of two of
these operators with opposite angles]. Here, we seek a transformation
of the following type:
\begin{equation}
|\psi_{\text{out}}\rangle = \hat{U}(\varphi) 
(\hat{N}_1\otimes\hat{N}_2)
\hat{U}(\varphi') |\psi_{\text{in}}\rangle := 
\hat{\not{\hspace*{-0.5ex}C}}
|\psi_{\text{in}}\rangle 
\end{equation}
with the two (conditional) nonlinear operators $\hat{N}_1$ and
$\hat{N}_2$.
A lengthy but straightforward calculation shows that the
operator sandwiched between the beam splitters does not have
tensor-product structure and thus cannot be regarded as single-mode
operators. In order to show that, we use a matrix technique. Let us
define a basis vector $|\textbf{e}\rangle$ as
\begin{equation}
|\textbf{e}^T\rangle =
(|00\rangle,|10\rangle,|01\rangle,|11\rangle,|20\rangle,|02\rangle) \,.
\end{equation}
Then, the input state $|\psi_{\text{in}}\rangle$ can be written as
$|\psi_{\text{in}}\rangle=\textbf{c}^T_{\text{in}}|\textbf{e}\rangle$.
In this basis the vector
$\textbf{c}_{\text{in}}^T=(c_{00},c_{10},c_{01},c_{11},0,0)$ transforms as
\begin{equation}
\label{eq:vectortransform}
\textbf{c}_{\text{out}} = \textbf{U}(\varphi')
(\textbf{N}_1\otimes \textbf{N}_2)
\textbf{U}(\varphi) \textbf{c}_{\text{in}}
\end{equation}
where the matrices $\textbf{U}(\varphi)$ etc. are the matrices
corresponding to the operators $\hat{U}(\varphi)$ etc. in the basis
$|\textbf{e}\rangle$ (these are not to be confused with the beam
splitter or transformation matrices used earlier on). For example, a
beam splitter is represented in this basis by the matrix
\begin{equation}
\textbf{U}(\varphi) = \left( \begin{array}{cccccc}
1&0&0&0&0&0 \\ 0&T&R&0&0&0 \\ 0&-R^\ast&T^\ast&0&0&0 \\
0&0&0&|T|^2-|R|^2&-\sqrt{2}R^\ast T&\sqrt{2}RT^\ast \\
0&0&0&\sqrt{2}RT&T&R^2 \\
0&0&0&-\sqrt{2}R^\ast T^\ast&R^{\ast 2}&T^{\ast 2}
\end{array} \right)
\end{equation}
with $|T|=\cos\varphi$ and $|R|=\sin\varphi$. The tensor product of
the two single-mode operators looks in this basis like
\begin{equation}
\label{eq:tensor}
\textbf{N}_1\otimes \textbf{N}_2 = \left( \begin{array}{ccc}
(N_1)_{00}(N_2)_{00}&(N_1)_{01}(N_2)_{00}&\cdots \\
(N_1)_{10}(N_2)_{00}&(N_1)_{11}(N_2)_{00}&\cdots \\
(N_1)_{00}(N_2)_{10}&(N_1)_{01}(N_2)_{10}&\cdots \\
\vdots & \vdots & \ddots 
\end{array} \right) \,.
\end{equation}
It is then relatively straightforward to show that there exists no
solution to Eq.~(\ref{eq:vectortransform}) with a matrix of the form
(\ref{eq:tensor}) that produces an output vector
$\textbf{c}_{\text{out}}^T=(c_{00},c_{10},c_{11},c_{01},0,0)$.

Therefore, in order to build a CNOT gate, we would have either to
refine our approach to include more general interferometric setups
(for which the original Knill-Laflamme-Milburn proposal is an example) 
or sandwich a controlled-$\hat{\sigma}_z$ gate between two Hadamard gates 
which we will show in the next section to be rather expensive.

\section{Crossing Fock layers}

Equipped with the knowledge about generating annihilation and creation
operators, we can start working on realisations of other operations that 
are harder to do but nevertheless needed to construct general quantum 
networks. By our Theorem, the `easy' operations are those that act within 
the same Fock layers. It is much harder to find suitable networks for 
operators that enable us to cross Fock layers \cite{ralph02a}. The
obvious choice  
consists of looking at single-qubit rotations first, i.e. the 
representations of the Pauli operators in the Fock basis,
\begin{eqnarray}
\hat{\sigma}_x &=& |0\rangle\langle 1|+|1\rangle\langle 0| \,,\\   
\hat{\sigma}_y &=& \frac{1}{i}(|0\rangle\langle 1|-|1\rangle\langle
0|) \,.
\end{eqnarray}
The construction of the corresponding photonic operators is almost   
obvious, once one takes care of the fact that one must not leave the 
Hilbert space of the qubits. Then it is clear that we have to choose
\begin{eqnarray}
\hat{\sigma}_x &=& \hat{a}-\hat{a}^\dagger(\hat{n}-1) \,,\\
\hat{\sigma}_y &=& \frac{1}{i}[\hat{a}+\hat{a}^\dagger(\hat{n}-1)]
\,.
\end{eqnarray}
In order to proceed further, we need a well-known result from
quantum-state engineering.

\vspace*{1ex}
\noindent
\textbf{Proposition 4}: Suppose one wants to generate the quantum state
\begin{equation}
|\psi_n\rangle = \sum\limits_{k=0}^n d_k |k\rangle
=\sum\limits_{k=0}^n \frac{d_k}{\sqrt{k!}} (\hat{a}^\dagger)^k
|0\rangle \,.
\end{equation}
Then one needs $n$ single-photon sources, at most $n$ coherent-state
sources, and at most $2n$ beam splitters and detectors.\\
\noindent
\textit{Proof}: The proof of this proposition follows closely the
result in \cite{ClausenArray} where it has been shown that the state
$|\psi_n\rangle$ can be generated by successive single-photon
additions and coherent shifts. The trick is to rewrite the state as
\begin{equation}
|\psi_n\rangle = \prod\limits_{k=1}^n
(\hat{a}^\dagger-\alpha_k^\ast)|0\rangle
\end{equation}
which is nothing but a decomposition of the polynomial in
$\hat{a}^\dagger$ into its root factors, where the $\alpha_k^\ast$ are
the roots of the polynomial.\proofend

Having generated the state $|\psi_n\rangle$, one can go ahead and
imprint it onto another state by mixing at a beam splitter. That leads
neatly to

\vspace*{1ex}   
\noindent
\textbf{Proposition 4a}: The polynomial
\begin{equation}
\hat{\cal P}_n=\sum\limits_{k=0}^n d_k (\hat{a}^\dagger)^k
\end{equation}
can be made to act upon a signal state by mixing the state
$\hat{\cal P}_n|0\rangle$ and the signal state at a single beam
splitter.

\noindent
\textit{Proof}: Let us assume that the signal state is again of the 
form
\begin{equation}
|\psi_{\text{in}}\rangle = \sum\limits_m \frac{c_m}{\sqrt{m!}}
(\hat{a}_1^\dagger)^m |0\rangle \,.
\end{equation}
Mixing $|\psi_{\text{in}}\rangle$ and $\hat{\cal P}_n|0\rangle$ at a
beam splitter, conditional on the second output being found in the  
vacuum state, we obtain after a short calculation
\begin{equation}
|\psi_{\text{out}}\rangle \propto \sum\limits_{k=0}^n d_k
\Lambda_{12}^k (\hat{a}_1^\dagger)^k \Lambda_{11}^{\hat{n}_1}
|\psi_{\text{in}}\rangle \,,
\end{equation}
from which we see that the coefficients have to be sufficiently
rescaled to achieve the desired goal.\proofend

In the same manner one can generate polynomials of annihilation
operators by projecting onto an engineered state. Combining both
processes opens up the opportunity to generate arbitrary polynomials
of creation and annihilation operators. However, this might not be the
best choice since doing quantum-state engineering of higher-order
polynomials is, as we have seen, an expensive task. Therefore, it   
might be advantageous to circumvent the problem of leaving the Fock   
layers of zero and one photon by projecting back onto this subspace
after performing a simplified version of the desired quantum
operation. For this, we introduce the KILL operator $\hat{K}$ as
\begin{equation}
\hat{K} = 1-\frac{1}{2}\hat{n}(\hat{n}-1)
\end{equation}
which, being a second-order polynomial in the number operator, requires 
two single-photon sources, two beam splitters and two
detectors. The Pauli operators can then be written as
\begin{eqnarray}
\hat{\sigma}_x &=& \hat{K} (\hat{a}+\hat{a}^\dagger) \,,\\
\hat{\sigma}_y &=& \hat{K} \frac{1}{i}(\hat{a}-\hat{a}^\dagger) \,.
\end{eqnarray}

With the theory presented above, we could go ahead and generate 
superposition states $|0\rangle+|1\rangle$ with the help of Proposition 
4a, superpose them onto the signal mode and perform a projection 
measurement onto a similar state. However, we will present a slightly 
different and more elegant method of achieving this purpose. Instead of 
preparing two copies of the superposition of vacuum and a single photon, 
we could prepare a Bell-type state $\sim|0,0\rangle+\lambda|1,1\rangle$ 
by the following method. Let us take a two-mode squeezed vacuum state 
of the form
\begin{equation}
|\mbox{TMSV}\rangle = \sqrt{1-q^2} \sum\limits_{n=0}^\infty q^n 
|n,n\rangle 
\end{equation}
and perform a Procrustean \cite{procrusten1,procrusten2} entanglement
concentration by acting on one  
mode of it with a first-order polynomial of the number operator as 
explained in the example (\ref{su(2)}). For appropriately chosen 
transmission coefficient $T$ of the beam splitter, we can generate in the 
limit $q\to 0$ the state
\begin{equation}
|\Phi(\lambda)\rangle = \frac{1}{\sqrt{1+|\lambda|^2}} \left[ 
|0,0\rangle+\lambda|1,1\rangle 
\right]
\end{equation}
to arbitrary accuracy in the trace-norm and for arbitrarily 
chosen $\lambda$ (details of this procedure can be found in 
\cite{Gaussification}). Using this state as the auxiliary-state source 
in an SU(3)-network that projects onto $|1,0\rangle$, we derive the 
following operation after applying the KILL operator:
\begin{equation}
c_0|0\rangle+c_1|1\rangle \to \Lambda_{21}c_1|0\rangle 
+\lambda\; \mbox{per }\bm{\Lambda}(3|1) c_0|1\rangle \,.
\end{equation}
Choosing $|\Lambda_{21}|=|\lambda\;\mbox{per }\bm{\Lambda}(3|1)|$ with an 
appropriate phase relation immediately leads to the desired Pauli operators.

A remark concerning the usage of continuous-variable states as resource is
of order here. In the described version of the Pauli operators we inject a
two-mode squeezed vacuum state into our network. This seems a simple and 
elegant method for getting  the desired result. In fact, we cannot see a
way around the usage of continuous-variable states at all, since even
for the creation of the superposition $|0\rangle+|1\rangle$, 
by Proposition~4, a coherent-state source is needed to displace the photon
creation operator $\hat{a}^\dagger$. A similar conclusion was reached by 
Lund and Ralph \cite{ralph02a}.

Another very important single-qubit operation is the Hadamard gate,
defined by
\begin{eqnarray}
|0\rangle &\to& \frac{1}{\sqrt{2}} \left( |0\rangle+|1\rangle
\right)\,,\\
|1\rangle &\to& \frac{1}{\sqrt{2}} \left( |0\rangle-|1\rangle \right)
\,.
\end{eqnarray}
This can also be written in operator form as
\begin{equation}
\hat{H}=\frac{1}{\sqrt{2}} \left( |0\rangle +(-1)^{\hat{n}}|1\rangle  
\right) \,,
\end{equation}
where the number operator is the one from the signal state! That is,
we swap signal and auxiliary states in the sense that we first produce a     
superposition of $|0\rangle$ and $|1\rangle$ and act conditionally on
it with the signal state. Effectively, the Hadamard gate becomes a  
(controlled) $\hat{\sigma}_z$ operation on the (auxiliary)
superposition state $(|0\rangle+|1\rangle)/\sqrt{2}$. In fact, one can
rewrite the operator $\hat{H}$ as
\begin{equation}
\label{eq:hadamard2}
\hat{H} = \frac{1}{\sqrt{2}} \left( |0\rangle +(1-2\hat{n}_1\hat{n}_2)
|1\rangle \right)
\end{equation}
which is effectively a two-mode operator. This is precisely the 
controlled-$\hat{\sigma}_z$ where we the second output is left 
unmeasured (sometimes called the DUMP `gate'). However, leaving
something unmeasured usually means to trace over the possible outcomes
which will destroy the purity and coherence of our desired
operation. The way around this problem is to act on the resulting
\textit{signal}-mode output with an operator $1+\hat{a}^\dagger$
(which can be prepared according to Proposition 4a) and then to
project onto the single-photon Fock state. 

Form this rather complicated construction we observe that the Hadamard
gate, and consequently also its multi-mode extension, the quantum
Fourier transform, are the hardest of all gates under investigation so
far. This result impacts the generation of gates that actually make
use of similar layer-crossings as the CNOT gate. For these type of
operations it seems that the constructive algorithm we have presented
in this article is not immediately applicable and this problem
requires further investigation.

%%%%%%%%%%%%%%%%%%%%%%%%%%%%%%%%%%%%%%%%%%%%%%%%%%%%%%%%%%%%%%%%%%%%%%
\section{Lossy beam splitters and non-perfect detectors}
\label{sec:lossy}

So far, we have restricted ourselves to perfect linear optics,
i.e. non-absorbing beam splitters and detectors with unit
efficiency. In practise, to achieve this situation is a hopeless
task. Instead, we have to make do with absorbing linear optical
elements and non-perfect detectors. What this amounts to in terms of
constructing our gates will be described in this section.

\subsection{Kraus decomposition}

We derive the Kraus decomposition of a lossy beam splitter. It is
known that an absorbing beam splitter represents a unitary evolution
in the extended Hilbert space of field and device modes. The unitary
operator can be written as \cite{Knoll99}
\begin{equation}
\hat{U}=\exp \left[ -i\left( \hat{\bm{\alpha}}^\dagger \right)^T \bm{\Phi}
\hat{\bm{\alpha}} \right] \,,
\end{equation}
where we use the notation
\begin{equation}
\hat{\bm{\alpha}} = {\hat{\bm{a}} \choose \hat{\bm{g}}} \,.
\end{equation}
Assume now the device to be initially in its vacuum state
$|0_3,0_4\rangle$. Then we can write the density operator of the
output field as
\begin{equation}
\hat{\varrho}_{\text{out}}^{(\text{F})} = \text{Tr}^{(\text{D})}
\left[ \hat{U} \left( \hat{\varrho}_{\text{in}}^{(\text{F})}
|0_3,0_4\rangle\langle 0_3,0_4| \right) \hat{U}^\dagger \right]
\end{equation}
and evaluate the trace in the coherent-state basis as
\begin{equation}
\hat{\varrho}_{\text{out}}^{(\text{F})} = \frac{1}{\pi^2} \int
d^2\alpha_3\,d^2\alpha_4\, \hat{E}_{\alpha_3,\alpha_4}
\hat{\varrho}_{\text{in}}^{(\text{F})} 
\hat{E}^\dagger_{\alpha_3,\alpha_4}
\end{equation}
where we have defined the Kraus operators
$\hat{E}_{\alpha_3,\alpha_4}$ as
\begin{equation}
\hat{E}_{\alpha_3,\alpha_4} = 
\langle \alpha_3,\alpha_4 | \hat{U} | 0_3,0_4 \rangle \,.
\end{equation}
They can be further simplified by using the relation \cite{Ma90}
\begin{equation}
e^{\hat{a}^\dagger M \hat{a}} = \sum\limits_{n=0}^\infty
\frac{:[\hat{a}^\dagger(e^M-1)\hat{a}]^n:}{n!}
\end{equation}
by writing
\begin{eqnarray}
\langle \alpha_3,\alpha_4 | \hat{U} | 0_3,0_4 \rangle &=&
\langle \alpha_3,\alpha_4| \exp \left[ -i\left(
\hat{\bm{\alpha}}^\dagger \right)^T \bm{\Phi} \hat{\bm{\alpha}}
\right]  | 0_3,0_4 \rangle
\nonumber \\ && \hspace*{-13ex}
= \langle \alpha_3,\alpha_4 | \sum\limits_{n=0}^\infty
\frac{:[\hat{\bm{\alpha}}^\dagger(\bm{\Lambda}-\openone)
\hat{\bm{\alpha}}]^n:}{n!}
| 0_3,0_4 \rangle
\nonumber \\ && \hspace*{-13ex}
= \sum\limits_{n=0}^\infty
\frac{:[\hat{\bm{a}}^\dagger(\textbf{T}-\openone)\hat{\bm{a}}
-\bm{\alpha}^+\textbf{S}\textbf{C}^{-1}\textbf{T}\hat{\bm{a}}]^n:}{n!}
e^{-\frac{1}{2}\bm{\alpha}^+\bm{\alpha}}
\nonumber \\ && \hspace*{-13ex}
= e^{-i\hat{\bm{a}}^\dagger\bm{\Phi}_T\hat{\bm{a}}}
e^{-\bm{\alpha}^+\bm{S}\bm{C}^{-1}\bm{T}\hat{\bm{a}}}
e^{-\frac{1}{2}\bm{\alpha}^+\bm{\alpha}} 
\end{eqnarray}
where we have used the definitions
\begin{eqnarray}
\bm{\Lambda} &=& \left( \begin{array}{cc}
\textbf{T} & \textbf{A} \\ -\textbf{S}\textbf{C}^{-1}\textbf{T} &
\textbf{C}\textbf{S}^{-1}\textbf{A} 
\end{array}\right)  =e^{-i\bm{\Phi}}\,,\\
\textbf{C} &=& \sqrt{\textbf{T}\textbf{T}^+} \,,\\
\textbf{S} &=& \sqrt{\textbf{A}\textbf{A}^+} \,,\\
\textbf{T} &=& e^{-i\bm{\Phi}_T}\,,\\
\hat{\bm{g}}|\alpha_3,\alpha_4\rangle &=&
\bm{\alpha}|\alpha_3,\alpha_4\rangle \,.
\end{eqnarray}
Therefore, we obtain the result that the Kraus operators for the
absorbing beam splitter are
\begin{equation}
\hat{E}_{\alpha_3,\alpha_4} =
e^{-i\hat{\bm{a}}^\dagger\bm{\Phi}_T\hat{\bm{a}}}
e^{-\bm{\alpha}^+\bm{S}\bm{C}^{-1}\bm{T}\hat{\bm{a}}}
e^{-\frac{1}{2}\bm{\alpha}^+\bm{\alpha}} \,.
\end{equation}
We can easily check that these operators become unitary when
absorption can be disregarded as $\textbf{T}$ becomes unitary (and
therefore $\bm{\Phi}_T$ hermitian), and $\textbf{S}$ vanishes. The
integration over $(\alpha_3,\alpha_4)$ can then be performed and gives
unity. What we also see is that these Kraus operators indeed
correspond to an absorption process for which the factor
$\exp[-\bm{\alpha}^+\textbf{S}\textbf{C}^{-1}\textbf{T}\hat{\bm{a}}]$ is
responsible. 

\subsection{Non-perfect detectors}

Second, we model a non-unit detector efficiency $\eta$ by replacing 
the projector $|n\rangle\langle n|$ by an appropriate POVM \cite{CondMeas}
\begin{equation}
\label{eq:povm}
|n\rangle\langle n| \to \hat{\Pi}(n) = \sum\limits_k {k \choose n}
\eta^n (1-\eta)^{k-n} |k\rangle\langle k| \,.
\end{equation}
This method does not take care of possible dark counts but reflects
the fact that direct photon counting may give values for the photon
number $n$ that actually came from higher Fock states $|k\rangle$, $k>n$.
This POVM is sometimes modelled by a perfect detector preceded by a
beam splitter with appropriately chosen transmissivity $|T|^2=\eta$.

\subsection{Example: a single beam splitter}

Let us consider a somewhat artificial example which nevertheless shows 
what happens when absorption and/or non-perfect detectors are present.
Suppose we were to implement the Pauli-$\hat{\sigma}_z$ gate with a
single beam splitter, a single-photon source and a single-photon
detector (note that this could have been done deterministically with a 
phase plate). We start off with a signal mode in a state
$c_0|0\rangle+c_1|1\rangle$ and mix it with a single photon. The
effect of the absorbing beam splitter is to produce a mixed state that
can be written in the form
\begin{equation}
\hat{\varrho}_{\text{out}}^{(\text{F})} =
|\psi_{\text{in}}(\textbf{T})\rangle\langle
\psi_{\text{in}}(\textbf{T})|  +
|\phi(\textbf{A})\rangle\langle \phi(\textbf{A})|
\end{equation}
where $|\psi_{\text{in}}(\textbf{T})\rangle$ is the state transformed
with the (non-unitary) transmission matrix $\textbf{T}$ and
$|\phi(\textbf{A})\rangle$ is a contribution that solely comes from
the absorption matrix $\textbf{A}$. We do not give the rather lengthy
expression here. Instead, we immediately give the result for the
non-normalised density matrix after applying the POVM (\ref{eq:povm}) as
\begin{eqnarray}
\label{eq:noisy_z}
\hat{\varrho}_{\text{out},1} &=& \eta |\psi_{\text{out}}\rangle\langle
\psi_{\text{out}} |
\nonumber \\ &&
+4\eta(1-\eta)|c_1|^2 |T_{12}|^2 |T_{22}|^2 |0\rangle\langle 0|
\nonumber \\ &&
+\eta |c_1|^2 \big( |T_{22}M_{11}+T_{12}M_{21}|^2 \nonumber \\ &&
+|T_{22}M_{12}+T_{12}M_{22}|^2  \big) |0\rangle\langle 0|
\end{eqnarray}
with the wanted output state
\begin{equation}
\label{eq:wanted}
|\psi_{\text{out}}\rangle = c_0 T_{22} |0\rangle
+c_1 (T_{11}T_{22}+T_{12}T_{21}) |1\rangle
\end{equation}
and the matrix $\textbf{M}=\textbf{S}\textbf{C}^{-1}\textbf{T}$.
Eq.~(\ref{eq:noisy_z}) has three parts: The first line is the wanted
outcome in which the transmission matrix can be chosen to give the
desired answer. The second line comes from the inefficient detector,
hence the POVM introduced in Eq.~(\ref{eq:povm}), whereas the last
two lines are the contributions due to the lossy beam splitter,
reflected in the appearance of the matrix $\textbf{M}$ that contains
the absorption matrix. The last expression can be simplified using the
fact that $\textbf{M}\textbf{M}^+=\openone-\textbf{T}\textbf{T}^+$ to
obtain 
\begin{eqnarray}
\hat{\varrho}_{\text{out},1} &=& \eta |\psi_{\text{out}}\rangle\langle
\psi_{\text{out}} |
\nonumber \\ &&
+4\eta(1-\eta)|c_1|^2 |T_{12}|^2 |T_{22}|^2 |0\rangle\langle 0|
\nonumber \\ &&
+\eta |c_1|^2 \big[ |T_{22}|^2+|T_{12}|^2-4|T_{12}|^2|T_{22}|^2
\nonumber \\ && \hspace*{7ex}
-|T_{11}T_{22}+T_{12}T_{21}|^2 \big] |0\rangle\langle 0| \,.
\end{eqnarray}
This expression shows that it is only necessary to know the
experimentally accessible transmission and reflection coefficients of
the beam splitter that make up the matrix $\textbf{T}$. Now we make
use of the fact that we actually wanted to generate a
Pauli-$\hat{\sigma}_z$ gate, meaning that we set in
Eq.~(\ref{eq:wanted}) $T_{11}T_{22}+T_{12}T_{21}=-T_{22}$. With that
we finally obtain for the (still unnormalised) output density matrix
\begin{eqnarray}
\hat{\varrho}_{\text{out},1} &=& \eta |T_{22}|^2 \hat{\sigma}_z
|\psi_{\text{in}}\rangle\langle \psi_{\text{in}} | \hat{\sigma}_z
\nonumber \\ &&
+4\eta(1-\eta)|c_1|^2 |T_{12}|^2 |T_{22}|^2 |0\rangle\langle 0|
\nonumber \\ && 
+\eta |c_1|^2 \left[ |T_{12}|^2-4|T_{12}|^2 |T_{22}|^2-3|T_{22}|^2 \right]
|0\rangle\langle 0| \,.\nonumber \\
\end{eqnarray}
The success probability for perfect operation is 
$p_{\text{success}}=|T_{22}|^2$.
A note of caution is appropriate here. Since we have fixed $T_{22}$
already, by reciprocity we have also fixed $T_{11}=T_{22}=T$. For
single-slab beam splitters that fixes $T_{12}=T_{21}=R$, too, so that
we are left with essentially a single number determining the fidelity
of our desired gate operation. To be more precise, note that
$|T|^2+|R|^2+|A|^2=1$ (setting 
$|A|^2=|A_{11}|^2+|A_{12}|^2=|A_{21}|^2+|A_{22}|^2$), and suppose that 
$T\in\mathbb{R}$. Then we immediately have that $R^2\in\mathbb{R}$, and 
choosing $\arg R=\pi/2$ we arrive at
\begin{equation}
T = \frac{\sqrt{3-2|A|^2}-1}{2} \,.
\end{equation}
With this choice for $T_{22}\equiv T$ we finally get
\begin{eqnarray}
\label{eq:final}
\hat{\varrho}_{\text{out},1} &=& \eta (2-|A|^2-\sqrt{3-2|A|^2}) 
\hat{\sigma}_z
|\psi_{\text{in}}\rangle\langle \psi_{\text{in}} | \hat{\sigma}_z
\nonumber \\ &&
+\eta(1-\eta)|c_1|^2 \left( |A|^4-3+2\sqrt{3-2|A|^2}\right)
|0\rangle\langle 0| \nonumber \\ && 
+\eta |c_1|^2 |A|^2(1-|A|^2) |0\rangle\langle 0|
\end{eqnarray}
which now only depends on two parameters, the absorption
coefficient $|A|$ of the beam splitter and the detector efficiency $\eta$. 
Again, the first line is the desired result, the second is due to the 
non-perfect detector, and the last line is the contribution of the 
absorption. Two special cases are notable here:
\begin{enumerate}
\item Without absorption ($|A|=0$), the third line in
Eq.~(\ref{eq:final}) vanishes and the numerical coefficient in the
second line takes the value of $2\sqrt{3}-3\approx 0.464$.
\item With perfect detectors ($\eta=1$), the second line vanishes and
we are left with a contribution $|A|^2(1-|A|^2)$ to the vacuum from
the last line.
\end{enumerate}
In principle, one could define a (state-dependent) gate fidelity or use
some more elaborate definition such as an average fidelity integrated over
all possible input states (with respect to some Haar measure) but this is
beyond the scope of this article.

%%%%%%%%%%%%%%%%%%%%%%%%%%%%%%%%%%%%%%%%%%%%%%%%%%%%%%%%%%%%%%%%%%%%%%
\section{Conclusions}
\label{sec:conclusions}

In this paper we have shown a constructive mechanism  for generating
arbitrary operators using only linear optics, single-photon sources,
and single-photon detectors. We have focused our attention primarily
on one-mode and two-mode situations, though the approach is easily extended
to multimode situations. We have shown what operations are easy and
what are potentially difficult. Operations that cause a change in the
Fock layers (for instance the Hadamard operator) are generally
difficult but not impossible. While the generation of the operators is
generally conditional on certain measurement results in the ancilla
modes, the operators can be made deterministic using various
teleportation protocols. Finally we hope this paper shows the power in
building the required operations from the fundamental resources rather
than fundamental gates. The SWAP operation illustrates this point
extremely well. From fundamental gates, three CNOTs are required to
build such an operation, however from fundamental resources, only two
beam splitters and a phase shifter are necessary. This approach open a
new way to think about operation generation.

\acknowledgments
We would like to thank Alexei Gilchrist for useful discussions as this
project developed. This work was funded in part by the Feodor--Lynen
program of the Alexander~von~Humboldt Foundation (SS), the United
Kingdom Engineering and Physical  Sciences Research Council and the
European Union projects RAMBOQ and QUIPROCONE.

%%%%%%%%%%%%%%%%%%%%%%%%%%%%%%%%%%%%%%%%%%%%%%%%%%%%%%%%%%%%%%%%%%%%%%
\appendix
\section{Permanents of unitary matrices}

Here we recall some elementary properties of permanents, mainly taken
from the only available monograph on this subject \cite{Minc}. 
The permanent of an $(n\!\times\!n)$-matrix $\textbf{A}$ is a
generalised matrix function, defined as 
\begin{equation}
\mbox{per }\textbf{A} = \sum\limits_{\{\sigma_i\} \in {\cal S}_n}
\prod\limits_{i=1}^n A_{i\sigma_i}
\end{equation}
where ${\cal S}_n$ is the symmetric group of cyclic permutations. Note
that the determinant of a matrix is similarly defined with the only
difference of a factor of $(-1)$ appearing in all terms depending on
the character (even or odd) of the permutation. The permanent of a
matrix generically appears in counting problems, i.e. combinatorics
and graph theory. In our case it is the probability amplitude of
detecting the state $|1\rangle^{\otimes N}$ after an input state of
the exactly the same form has been transformed by an
SU($N$)-network. In that sense, it naturally appears here as well
since the combinatorial problem is here to (re-)distribute $N$ single
photons among $N$ single-photon detectors.

The Marcus--Newman theorem states that the following inequality holds
for all $(m\!\times\!n)$-matrices $\textbf{A}$ and
$(n\!\times\!m)$-matrices $\textbf{B}$: 
\begin{equation}
|\mbox{per }\textbf{AB}|^2 \le \mbox{per }\textbf{AA}^\ast
\mbox{per }\textbf{BB}^\ast \,. 
\end{equation}
An immediate consequence is that (setting $\textbf{B}\!=\!\openone$),
if $\textbf{U}$ is unitary, then
\begin{equation}
\label{eq:a1}
|\mbox{per }\textbf{U}| \le 1 \,.
\end{equation}
Note that this condition also follows immediately from the
probabilistic interpretation given above. Eq.~(\ref{eq:a1}) tells us
that the range of the permanent of a unitary matrix lies in the unit
disk in the complex plane. In fact, the same conclusion can be drawn
for the permanents of principal submatrices of unitary matrices by
recalling that a unitary matrix consists of rows (or columns) of
orthogonal unit vectors. For example, let us consider
$\mbox{per }\bm{\Lambda}(1|1)$ of $\bm{\Lambda}\!\in$ SU(3). We have
\begin{equation}
|\mbox{per }\bm{\Lambda}(1|1)| =
|\Lambda_{22}\Lambda_{33}+\Lambda_{23}\Lambda_{32}| \,.
\end{equation}
Since $|\Lambda_{23}|\le\sqrt{1-|\Lambda_{22}|^2}$ and
$|\Lambda_{32}|\le\sqrt{1-|\Lambda_{33}|^2}$, we know that
\begin{eqnarray}
|\mbox{per }\bm{\Lambda}(1|1)| &\le&
|\Lambda_{22}\Lambda_{33}|+|\sqrt{(1-|\Lambda_{22}|^2)(1-|\Lambda_{33}|^2)}|
\nonumber \\ &=& |\cos\varphi\cos\Theta|+|\sin\varphi\sin\Theta|
\nonumber \\ &=& |\cos(\varphi\pm\Theta)| \le 1 \,.
\end{eqnarray}
Similar relations hold for $\mbox{per }\bm{\Lambda}(2|2)$ and
$\mbox{per }\bm{\Lambda}(3|3)$ and indeed for all permanents of submatrices
of unitary matrices.

%%%%%%%%%%%%%%%%%%%%%%%%%%%%%%%%%%%%%%%%%%%%%%%%%%%%%%%%%%%%%%%%%%%%%%


\begin{thebibliography}{99}

\bibitem{dowling02} 
J.P.~Dowling, G.J.~Milburn, 
\textit{Quantum Technology: The Second Quantum Revolution},
\textit{quant-ph/0206091}.

\bibitem{feymann82}
R.P.~Feymann, Int. J. Theor. Phys. \textbf{21}, 467 (1982).

\bibitem{deutsch85}
D.~Deutsch, Proc. R. Soc. London. A \textbf{400}, 97 (1985); 
Proc. R. Soc. London. A \textbf{425}, 73 (1989).

\bibitem{shor95} P.~Shor, In Proceedings 35th Annual Symposium on
Fundamentals of Computer Science, IEEE Press, Los Alamitos, CA, 1994;
Phys. Rev. A \textbf{52}, 2493 (1995).

\bibitem{machines} See for instance: H.-K.~Lo, S.~Popescu and
T.P.~Spiller (eds.),  
\textit{Introduction to Quantum Computation and Information}, (World
Scientific Publishing, 1998); 
Fortschr. Phys.  48, Number 9-11, Special Focus Issue: "Experimental
Proposals for Quantum Computers", eds. S.~Braunstein and H.-K.~Lo (2000); 
R.G.~Clark (ed.),
\textit{Experimental Implementation of Quantum Computation}, (Rinton
Press, 2001). 

\bibitem{NielsenChuang}
M.A.~Nielsen and I.L.~Chuang, \textit{Quantum Computation and Quantum
Information} (Cambridge University Press, Cambridge, 2000).


\bibitem{milburn88}
G.J.~Milburn, Phys. Rev. Lett.{ \bf 62}, 2124 (1988).

\bibitem{shen84}
Y.R.~Shen, \textit{The Principles of Nonlinear Optics} (Wiley, New
York, 1984). 

\bibitem{KLM}
E.~Knill, R.~Laflamme and G.~J.~Milburn, Nature, {\bf 409}, 46 (2001).

\bibitem{footnote-1}
Several schemes using non-optical elements have 
also been proposed for producing the non-linear phase shifts. 
They can have the advantage of performing the above transformation
nearly deterministically.  
See for instance  A.~Gilchrist and G.J.~Milburn,
\textit{quant-ph/0208157}. 

\bibitem{ralph02a} 
A.P.~Lund and T.C.~Ralph, Phys. Rev. A {\bf 66}, 032307 (2002).

\bibitem{Ralph01}
T.C.~Ralph, A.G.~White, W.J.~Munro and G.J.~Milburn,
Phys. Rev. A {\bf 65}, 012314 (2001).

\bibitem{kok02}
P.~Kok, H.~Lee, and J.P.~Dowling,
Phys. Rev. A \textbf{66}, 063814 (2002).

\bibitem{others}
T.B.~Pittman, B.C.~Jacobs and J.D.~Franson, Phys. Rev. A {\bf 64},
062311 (2001);
T.~Rudolph and J.-W.~Pan, \textit{A simple gate for linear optics
quantum computing}, \textit{quant-ph/0108056};
T.C.~Ralph, N.K.~Langford, T.B.~Bell, and A.G.~White,
Phys. Rev. A {\bf 65}, 062324 (2002);
A.~Gilchrist, W.J.~Munro, and A.G.~White, Phys. Rev. {\bf A} 67,
040304 (2003);
M.~Koashi, T.~Yamamoto and N.~Imoto, \pra {\bf 63},
030301(R) (2001);
E.~Knill, \textit{A Note on Linear Optics Gates by Post-Selection},
\textit{quant-ph/0110144};


%\bibitem{Ralph01}
%T.C.~Ralph, A.G.~White, W.J.~Munro and G.J.~Milburn,
%Phys. Rev. A {\bf 65}, 012314 (2001).

%\bibitem{pittman01}
%T.B.~Pittman, B.C.~Jacobs and J.D.~Franson, Phys. Rev. A {\bf 64},
%062311 (2001).

%\bibitem{rudolph01} 
%T.~Rudolph and J.-W.~Pan, \textit{A simple gate for linear optics
%quantum computing}, \textit{quant-ph/0108056}. 

%\bibitem{ralph02} 
%T.C.~Ralph, N.K.~Langford, T.B.~Bell, and A.G.~White,
%Phys. Rev. A {\bf 65}, 062324 (2002).

%\bibitem{kok02}
%P.~Kok, H.~Lee, and J.P.~Dowling,
%Phys. Rev. A \textbf{66}, 063814 (2002).

%\bibitem{munro03}
%A.~Gilchrist, W.J.~Munro, and A.G.~White, Phys. Rev. {\bf A} 67,
%040304 (2003). 

%\bibitem{koasho01}
%M.~Koashi, T.~Yamamoto and N.~Imoto, \pra {\bf 63},
%030301(R) (2001). 

%\bibitem{knill01}
%E.~Knill, \textit{A Note on Linear Optics Gates by Post-Selection},
%\textit{quant-ph/0110144}. 

\bibitem{gottesman99}
D.~Gottesman, and I.L.~Chuang, Nature, { \bf 402},
390--393 (1999).

\bibitem{pittman02}
T.B.~Pittman, B.C.~Jacobs, and J.D.~Franson, Phys. Rev. Lett. {\bf 88},
257902 (2002). 

\bibitem{pittman02a}
T.B.~Pittman, B.C.~Jacobs, and J.D.~Franson, Phys. Rev. A {\bf 66},
052305 (2002). 

\bibitem{pittman03}
T.B.~Pittman, M.J.~Fitch, B.C.~Jacobs, and J.D.~Franson, 
\textit{Experimental Controlled-NOT Logic Gate for Single Photons},
\textit{quant-ph/0303095}.
 
\bibitem{kok02a}
P.~Kok, C.P.~Williams, and J.P.~Dowling, 
\textit{Practical quantum repeaters with linear optics and double-photon guns},
\textit{quant-ph/0203134}.

\bibitem{beamsplitter}
B.~Yurke, S.L.~McCall, and J.R.~Klauder, Phys. Rev. A \textbf{33},
4033 (1986);
S.~Prasad, M.O.~Scully, and W.~Martienssen, Opt. Commun. \textbf{62},
139 (1987);
Z.Y.~Ou, C.K.~Hong, and L.~Mandel, Opt. Commun. \textbf{63}, 118
(1987);
H.~Fearn and R.~Loudon, Opt. Commun. \textbf{64}, 485 (1987);
M.A.~Campos, B.E.A.~Saleh, and M.C.~Teich, Phys. Rev. A \textbf{40},
1371 (1989);
U.~Leonhardt, Phys. Rev. A \textbf{48}, 3265 (1993).


\bibitem{plk1}
A.K.~Ekert and P.L.~Knight, Phys. Rev. A {\bf 42}, 487 (1990);
A.K.~Ekert and P.L.~Knight, Phys. Rev. A {\bf 43}, 3934 (1991).

\bibitem{stenholm1}
P.~T\"{o}rm\"{a} and S.~Stenholm, Phys. Rev. A {\bf 54}, 4701 (1996).

\bibitem{Reck94}
M.~Reck, A.~Zeilinger, H.J.~Bernstein, and P.~Bertani,
Phys. Rev. Lett. \textbf{73}, 58 (1994).

\bibitem{VogelWelsch}
W.~Vogel, S.~Wallentowitz, and D.-G.~Welsch, \textit{Quantum Optics: An 
Introduction} (Wiley-VCH, Berlin, 2001).

\bibitem{Louisell} W. Louisell, \textit{Quantum Statistical Properties of 
Radiation} (Wiley, New York, 1974).

\bibitem{CondMeas}
J.~Clausen, M.~Dakna, L.~Kn\"oll, and D.-G.~Welsch,
J. Opt. B: Quantum Semiclass. Opt. \textbf{1}, 332 (1999).

\bibitem{Lvovsky}
A.I.~Lvovsky and J.~Mlynek, Phys. Rev. Lett. \textbf{88}, 250401 (2002).

\bibitem{footnote-2}
The last term in Eq.~(\ref{eq:prop3}) is in fact just the
definition of the permanent of the principal submatrix $\bm{\Lambda}(1|1)$.

\bibitem{Minc}
H.~Minc, \textit{Permanents} (Addison--Wesley, London, 1978).


\bibitem{vacuum-detector}
A vacuum detector is a device that is about to distinguish the vacuum
state from any other Fock state. Mathematically it can be represented
by the projector $\Pi_0=|0\rangle \langle 0|$. 


\bibitem{ClausenArray}
J.~Clausen, M.~Dakna, L.~Kn\"oll, and D.-G.~Welsch,
Acta Phys. Slov. \textbf{49}, 653 (1999).


\bibitem{procrusten1}
C.H.~Bennett, H.J.~Bernstein, S.~Popescu, and
B.~Schumacher, Phys. Rev. A {\bf 53}, 2046 (1996).

\bibitem{procrusten2}
R.T.~Thew and W.J.~Munro, Phys. Rev. A {\bf 63}, 30302R (2001); 
Phys. Rev. A \textbf{64}, 022320 (2001).

\bibitem{Gaussification}
D.E.~Browne, J.~Eisert, S.~Scheel, and M.B.~Plenio, 
\textit{quant-ph/0211173}, to be published in Phys. Rev. A.

\bibitem{Knoll99}
L.~Kn\"oll, S.~Scheel, and D.-G.~Welsch,
\textit{QED in dispersing and absorbing media}, in
\textit{Coherence and Statistics of Photons and Atoms} ed. by
J.~Pe\v{r}ina (Wiley, New York, 2001);
L.~Kn\"oll, S.~Scheel, E.~Schmidt, D.-G.~Welsch, and A.V.~Chizhov,
Phys. Rev. A \textbf{59}, 4716 (1999).

\bibitem{Ma90}
X.~Ma and W.~Rhodes, Phys. Rev. A \textbf{41}, 4625 (1990).


\end{thebibliography}
\end{document}